\documentclass{emulateapj}

\newcommand{\LT}{L-~$\&$~T-}

\newcommand{\myemail}{Russell.Ryanjr@asu.edu}
\newcommand{\super}[1]{$^{#1}$}

\newcommand{\tab}[1]{Table~\ref{#1}}
\newcommand{\fig}[1]{Figure~\ref{#1}}

\newcommand{\lon}{l^{{\rm II}}}
\newcommand{\lat}{b^{{\rm II}}}

\newcommand{\nfields}{\ensuremath{15}}
\newcommand{\nltds}{\ensuremath{28}}
\newcommand{\minz}{\ensuremath{350}}
\newcommand{\errz}{\ensuremath{50}}
\newcommand{\finz}{\ensuremath{\minz\!\pm\!\errz}}
\newcommand{\rscl}{\ensuremath{2100}}
\newcommand{\blim}{\ensuremath{15}}
\newcommand{\absb}{\ensuremath{|\lat|\!\lesssim\!\blim\degr}}
\newcommand{\iz}{\ensuremath{(i'\!-\!z')}}
\newcommand{\gi}{\ensuremath{(g'\!-\!i')}}
\newcommand{\zj}{\ensuremath{(z'-J)}}

\shorttitle{Galactic \LT Dwarfs}
\shortauthors{Ryan et al.}

\begin{document}
\title{Constraining   the   Distribution   of   \LT  Dwarfs   in   the
Galaxy\footnote{Based  on observations made  with the  NASA/ESA Hubble
Space Telescope, obtained from the Data Archive at the Space Telescope
Science   Institute,  which   is  operated   by  the   Association  of
Universities for Research in  Astronomy, Inc., under NASA contract NAS
5-26555.}}

\author{R. E. Ryan Jr., N. P. Hathi, S. H. Cohen \& R. A. Windhorst\altaffilmark{2}}
\affil{Department of Physics and Astronomy, Arizona State University, Tempe, AZ 85281}
\email{\myemail}

\begin{abstract}

We estimate the  thin disk scale height of  the Galactic population of
\LT dwarfs based  on star counts from $\nfields$  deep parallel fields
from  the Hubble Space  Telescope.  From  these observations,  we have
identified $\nltds$ candidate \LT dwarfs based on their \iz\ color and
morphology.   By comparing  these  star counts  to  a simple  Galactic
model,  we  estimate  the  scale  height  to  be  $\finz$~pc  that  is
consistent with the increase in vertical scale with decreasing stellar
mass  and is  independent  of reddening,  color-magnitude limits,  and
other Galactic parameters.  With this refined measure, we predict that
less than $10^{9}$~M$_{\odot}$ of the Milky Way can be in the form \LT
dwarfs, and confirm  that high-latitude, $z\!\simeq\!6$ galaxy surveys
which use  the $i'$-band  dropout technique are  97-100\% free  of \LT
dwarf interlopers.

\end{abstract}

\keywords{stars: low-mass, brown dwarfs --- Galaxy: structure ---  Galaxy: stellar content}

\section{Introduction} \label{introduction}

The method of counting stars to infer the shape and size of the Galaxy
is as old as astronomy  itself.  Earliest efforts using this technique
were  famously flawed  as they  often relied  on insufficient  data or
wholly incorrect assumptions \citep[eg.][]{herschel,kapteyn22}.  Aided
by  advanced  technology, \citet[B\&S;][]{bahc80,bahc81}  demonstrated
that the true power of star  counts is realized when they are compared
to  simulations  of the  fundamental  equation  of stellar  statistics
\citep{vonsee}.   The  B\&S  method  relies  heavily  on  the  assumed
luminosity  functions  and  density  distributions,  and  has  been  a
standard method for many subsequent studies.

The  Galaxy is  traditionally characterized  by having  a Population~I
disk  and Population~II  spheroid.   In  a series  of  studies of  the
exponential disk, \citet{gil83} and \citet{gil84} established the need
for  a  thick  and thin  disk  whose  scale heights  are  inversely
proportional  to the  masses of  the studied  stars  \citep[see Table~1
in][]{siegel02}.  The  standard description of the Galactic  halo is a
de Vaucouleur  or power-law  profile, while the Besan\c{c}on flattened
spheroid                  with                 $c/a\!\approx\!0.5-0.8$
\citep{bahc84,robin00,larsen03,robin03}  being  the currently  favored
parameters.  Thorough  discussions of star counts  and their relevance
to   Galactic  structure   are  given   in  the   Annual   Reviews  by
\citet{bahc86}, \citet{gil89}, and \citet{majewski}.

Many of the Galactic models and the majority of the literature examine
on  relatively luminous dwarf  and/or giant  stars and  rarely address
sub-stellar  objects.    The  discovery  of   the  first  extra-solar,
sub-stellar  object,  Gliese~229B  \citep{naka95,opp95} motivated  the
creation of the  L and T spectral classes.   With surface temperatures
ranging from 750--2200~K \citep{burg99}, the \LT dwarfs can contaminate
searches for  $z\!\simeq\!6$, $i'$-band dropout  objects \citep{yan03}
by  mimicking the extremely  red broad-band  colors.  This  effect has
remained largely unquantified due to insufficient knowledge of the \LT
dwarf IMF, Galactic distribution,  and local number density.  Previous
work     on     their     IMF     and     local     number     density
\citep{reid99,chab01,chab02,liu02}    has   suffered    from   limited
statistics.  With the deep imaging of \nfields\ Hubble Space Telescope
(HST) Advanced  Camera for Surveys  (ACS) parallel fields,  this study
increases the number  of {\it faint} dwarfs by  adding $\nltds$ to the
growing list.   Little work has been  done on the scale  height of \LT
dwarfs;  \citet{liu02}  and  \citet{grapes} estimated  100--400~pc  as
based  on  a single  object  or  a 3--4  objects  in  a single  field.
Therefore  the primary  goal of  this work  is to  estimate  the scale
height of the \LT dwarf  population by comparing the surface densities
from the ACS parallel fields to the Galactic structure models.

\section{Observations} \label{observations}
The  \LT  dwarf  candidates  were  selected  from  $\nfields$  HST/ACS
parallel  fields covering  a  broad range  in  Galactic latitudes  and
longitudes (see  \tab{thetable}).  All ACS fields have  at least three
independent exposures in F775W (SDSS-$i'$) and F850LP (SDSS-$z'$) with
a  total exposure  time  of  2--10~ks per  bandpass.   All fields  are
$\geq$90\% complete  at $z'$(AB){$\simeq$}26.0~mag \citep{yan04a}.  We
adopt the AB magnitude system \citep{oke83}.

After combining the individual ACS  frames into final stacks using the
PyRAF-based  script {\it  multidrizzle} \citep{multi},  the SExtractor
package  \citep{sex} was  used  in double-input  mode  to perform  the
matched-aperture photometry.  The F850LP  stack was used to define the
optimal  apertures for  the  flux measurements  in  both stacks.   For
source detection, we used  a 5$\times$5 Gaussian smoothing kernel with
a FWHM of  2.0 pixels, which is approximately the same  as the FWHM of
the ACS point-spread function (PSF) on both image stacks.  
We used total magnitudes (corresponding to the {\it MAG\_AUTO} option
in  SExtractor)  for  the  photometry  and  adopted  the  zero  points
published in HST ACS Instrument Science Report \citep{HSTzero}.

\begin{figure}
\epsscale{.85}
\plotone{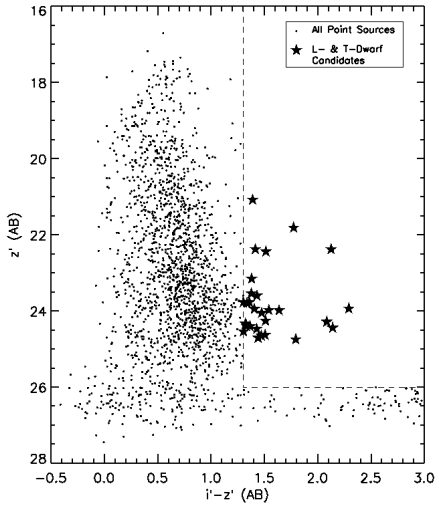}
\caption{Color-magnitude diagram for all point sources detected in our
$\nfields$~ACS fields.  The small  dots represent all objects that met
the  stellar  morphology  classification,  the  solid  stars  are  the
$\nltds$  \LT dwarf  candidates, and  the dashed  lines  represent the
imposed  color-magnitude limits.  Many  point sources  associated with
diffraction  spikes,  field edges,  and  spurious detections  (objects
smaller  than the  PSF,  likely residual  cosmic  rays) brighter  than
$z'\!=\!26$~mag  were manually  removed.  The  similar  points fainter
than $z'\!=\!26$~mag were not removed from this figure.  The sample of
stellar        candidates,        becomes        incomplete        for
$z'\!\geq\!25$~mag.}\label{colormag}
\end{figure}

\begin{figure}
\epsscale{1.0}
\plotone{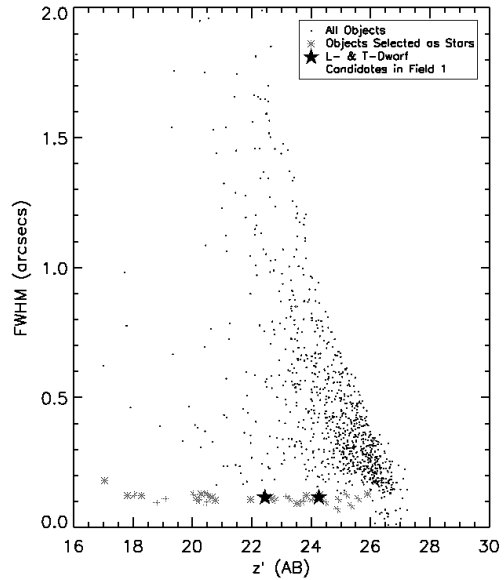}
\caption{The observed FWHM for all  objects in Field~1 is plotted as a
function of  apparent $z'$ magnitude.  The small  dots, asterisks, and
filled stars represent  all objects, the point sources  used to define
the  ``stellar-locus,''  and the  objects  selected  as candidate  \LT
dwarfs.   Clearly  many  objects  were recognized  as  point  sources,
however only two met the \iz\ color criterion.}\label{elephant}
\end{figure}

As   a  cursory   selection,  all   objects  with   \iz$>$1.3~mag  and
$z'\!<$26.0~mag    were    considered.     \fig{colormag}    is    the
color-magnitude  diagram  for  all   point  sources  detected  in  all
$\nfields$~fields.  Dashed lines  indicate the imposed color-magnitude
limits and the large, filled stars represent the candidate \LT dwarfs.
For  the  five  fields  for  which  the  F475W  (SDSS-$g'$)  band  was
available, we  required candidates to have  \gi$>$0~mag.  Objects near
the detector edges were not  considered, yielding an effective area of
$\sim$9~arcmin$^2$  per ACS field.   Extended objects  were eliminated
from  the analysis  by using  the  FWHM parameter  of SExtractor.   In
\fig{elephant}, we plot  the FWHM as a function  of apparent magnitude
for each  object in  Field~1 as  small dots.  The  locus of  points at
FWHM$\sim\!0\farcs13$  and  $z'\!<\!26$~mag  (hereafter the  ``stellar
locus''), represented  as asterisks,  are the unresolved  objects with
the minimum  possible FWHM.  In addition to  the above color-magnitude
criterion, we  required all  \LT dwarf candidates  to lie  within this
locus.  Sources  of contamination to  these criteria are  discussed in
\S~\ref{contamination}.

\begin{table}
\caption{\LT Dwarf Number Counts}
\label{thetable}
\begin{tabular*}{0.48\textwidth}
   {@{\extracolsep{\fill}}crrc}
\hline
\hline
\multicolumn{1}{c}{Field}&\multicolumn{1}{c}{$\lon$}&\multicolumn{1}{c}{$\lat$}&\multicolumn{1}{c}{Number}\\
\multicolumn{1}{c}{No.}&\multicolumn{1}{c}{(deg)}&\multicolumn{1}{c}{(deg)}&\multicolumn{1}{c}{(per 9 arcmin$^{2}$)}\\
$ $&$ $&$ $&$ $\\
\hline
1&115.018&$+$46.681&2\\
2&164.056&$-$75.750&1\\
3&169.188&$-$59.664&0\\
4&279.934&$-$19.990&3\\
5&165.876&$+$36.396&0\\
6&280.782&$+$68.293&0\\
7&113.107&$+$28.548&2\\
8&293.996&$-$41.466&0\\
9&316.829&$-$40.490&1\\
10&105.103&$+$7.075&11\\
11&298.138&$-$13.885&6\\
12&92.666&$+$46.378&1\\
13&70.106&$+$62.876&0\\
14&251.327&$-$41.444&0\\
15&216.142&$+$54.561&1\\
$ $&$ $&$ $&$ $\\
\hline
\end{tabular*}
\end{table}

Typically Galactic structure studies examine star counts
from   one  or  many   shallow  fields   with  large   surveyed  areas
\citep[eg.][]{siegel02,larsen03}.    Thus  the   data  of   the  2MASS
\citep{burg99,kirk99},     DENIS    \citep{delf99},     and/or    SDSS
\citep{stra99,tsve00,hawl02} are natural choices to study the Galactic
distribution of the \LT dwarfs.  These surveys have significantly more
detection area than our HST dataset and their \LT dwarfs are typically
closer  to  the  Sun  than  $\sim\!300$~pc  or  1~disk  scale  height.
Therefore  to  avoid {\it  extrapolating}  the  vertical scale  height
beyond this  distance, we chiefly  analyzed the HST dataset  where all
sample stars are likely more distant than 1~$e$-folding length.

\section{The Simple Galactic Model}\label{model}

The   Galactic   structure    models   were   made   by   distributing
10\super{10}~points  according  to   an  exponential  disk that was 
motivated by the light profiles of edge-on galaxies \citep{degrijs97}:
$n(r,\theta,z)=n_0\,\exp{\left(\frac{R_0-r}{r_l}\right)}\exp{\left(\frac{Z_0-z}{z_h}\right)}$,
where   $r_l$=$\rscl$~pc  is   the  radial   scale  length   found  by
\citet{porc98},  $n_0$=0.12~pc$^{-3}$ is the  local space  density \LT
dwarfs  taken  from \citet{chab02},  and  $R_0$=8~kpc and  $Z_0$=15~pc
\citep{yama92}  are the  solar radius  and height,  respectively.  The
vertical  scale height,  $z_h$, is  the  free parameter  and found  by
minimizing the  squared difference between the number  counts from the
model and the  HST data.  Altering the assumed  coordinates of the Sun
and the radial  scale length have little effect  on the vertical scale
height  estimate.   To generate absolute magnitudes, we adopted the 
$J$-band luminosity function of \citet{cruz03} and the \zj\ colors of 
\citet{hawl02} over the appropriate range of spectral type.

\begin{figure}
\epsscale{1.0}
\plotone{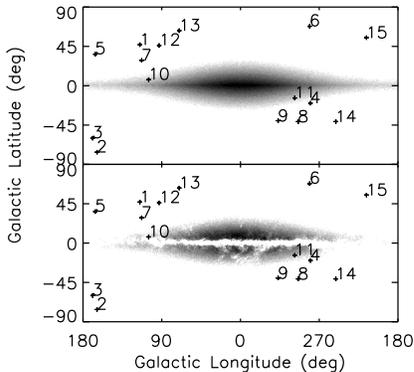}
\caption{A sample realization of the Monte Carlo simulation   
with $10^{10}$ random points with the best-fit vertical scale  height  of 
\minz\ pc.  The over-plotted crosses  with numbers represent  each of the  
observed HST/ACS field in  \tab{thetable}.  The effects of the 
\citet{schl98} dust maps are readily apparent when comparing the upper (no 
extinction) and lower (extinction) panels.}\label{modelgal}
\end{figure}

The effects  of interstellar extinction  were included into  the model
using the  COBE dust  maps of \citet{schl98}  in two ways.   First, we
assumed that  each point was  located beyond the dust,  establishing a
lower bound  on the  model counts.  Alternatively,  an upper  bound is
reached  by assuming  that the  Galaxy has  {\it no}  dust whatsoever.
Since an overwhelming majority of the dust is localized to \absb\ (see
\fig{modelgal}) and  only two of our  observed ACS fields  are in this
range, either approach yielded the same result within
the  uncertainties,   therefore  we  adopted  the   third  method  for
simplicity.  \fig{modelgal} is  a representative realization the model
with a scale height of \minz\ pc, the with locations of the $\nfields$
observed fields are indicated with plus signs.

The  canonical disk/spheroid Galaxy  likely has  additional components
\citep{bahc86}, the models used  here did not contain any contribution
from the  Galactic bulge  or a two-component  disk \citep{gil83,gil84}
for  the  following reasons.   First,  a  bulge  distribution was  not
modeled     since     its      radius     is     $\sim$1--2~kpc     or
$\sim$7$\degr$--14$\degr$, and every field is well beyond 14$\degr$ of
the Galactic center, hence we do  not require a bulge component in the
model.  Second,  this sample contains  only $\nltds$ \LT  dwarfs which
are likely within $\sim$1000~pc  (based on the luminosity function) of
the Sun.  Since the thick disk has a scale height of $\gtrsim$1000~pc,
we  expect the  star counts  to be  dominated by  a {\it  single} disk
population.  Moreover,  With only  $\nltds$ candidates the  models and
analyses must remain simple and straightforward.

\section{Analysis} \label{analysis}

\begin{figure}
\epsscale{1.0}
\plotone{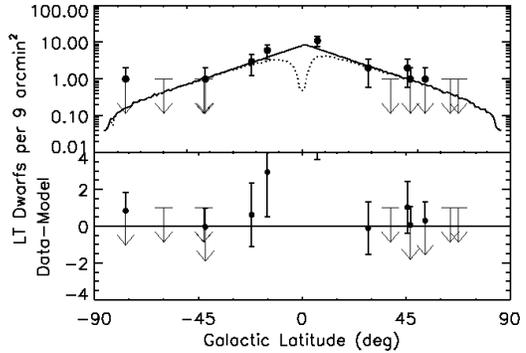}
\caption{{\bf  Top panel:}  model  surface density  as  a function  of
Galactic latitude.  
Here  we have averaged over all  longitudes with a
2.5$\sigma$-clipping to  better handle  the sparse statistics  at high
latitudes.
The dotted and full lines indicate models with and without
the   extinction  corrections, respectively.  
The data from the  \nfields\ ACS fields are plotted for
comparison as  filled circles  and downward arrows  as an  upper limit
when either zero or one  object was detected.  {\bf Bottom panel:} the
residuals from  the upper  panel as a  function of  Galactic latitude.
Clearly the data where the  dust corrections are large (ie. \absb) are
the most deviant.  The model used  in both panels has a vertical scale
height  of \minz\ pc.   The two  fields outlying  fields at  \absb\ are
discussed in \S~\ref{contamination}.  }\label{surfres}
\end{figure}

Despite this work utilizing the largest dataset of \LT dwarfs compiled
from HST observations, the star  counts remain very sparse requiring a
simple  analysis  scheme.   Using  the  grid  of  Monte  Carlo  models
described  in  \S~\ref{model},  we   sought  the  scale  height  which
minimizes the squared difference between the integrated star counts of
the model and those from the HST/ACS dataset.  For the fields where no
\LT  dwarf candidates were  found, we  assumed an  upper limit  of one
object  (per  field)  could   have  been  detected  and  perform  this
minimization technique simultaneously  on all $\nfields$ fields.  This
procedure yielded a  vertical scale height of \finz~pc.   In the upper
panel of  \fig{surfres}, we plot the modeled  surface density averaged
over Galactic longitude as a function of Galactic latitude as computed
from the  model with  a scale  height of \minz~pc,  with the  HST data
points from \tab{thetable} over-plotted for comparison.  The residuals
in the  lower panel  clearly demonstrate that  the model with  a scale
height  of  \minz~pc   reproduces  the HST  star  counts  for
$\left|\lat\right|\!\geq\!15\degr$ where dust extinction is minimal.

\subsection{Sources of Contamination} \label{contamination}
While  all \LT dwarf  candidates  were systematically  found by
color  and FWHM  criteria,  each  was visually  confirmed  as a  point
source.  However,  the  color-magnitude rules  outlined  in
\S~\ref{observations} potentially find  three classes of contaminates:

(1) The primary motivation  of  this  study is  to  reliably correct  the
$z\!\simeq\!6$, $i'$-band  dropout galaxy surveys  for interloping \LT
dwarfs \citep{yan02,yan03,yan04a,yan04b}.  Since our method is similar
to the  $i'$-band dropout technique, we  expect possible contamination
to  the star  counts from  the $z\!\simeq\!6$  galaxies.  In  a recent
study of  the Hubble Ultra-Deep Field (HUDF),  \citet{yan04b} find 108
$i'$-band  dropouts using  the \iz$>$1.3~mag  color  selection.  Their
sample has a median $z'$-band magnitude of 28.5~mag, and contains only
three objects brighter than our limit of $z'\!=$26.0~mag.  Each of
these three objects is considerably extended and could not be mistaken
for a  point-source.  Since the majority  of unresolved $z\!\simeq\!6$
galaxies  will   be  $\sim$2.5~mag   too  faint,  we   conclude  their
contamination in our sample must be negligible.

(2) Another known  source of  possible contaminates comes  from dusty,
elliptical  galaxies with  redshifts  $1.0\!\leq\!z\!\leq\!1.5$, whose
4000~$\mbox{\AA}$  break  occurs  between  the  $i'$  and  $z'$  bands
\citep{yan03}.   With a  typical  color of  \iz$\sim$1.0~mag, many  of
these objects would appear too blue in the absence of extreme internal
reddening.    Moreover  the   visual   identification  confirms   only
point-like objects populate the ``stellar locus'' in the lower left of
\fig{elephant}.  Hence, extended elliptical galaxies could not grossly
corrupt our sample.

(3) In  addition to  the above  extragalactic sources,  we anticipated
contamination  from  galactic  M-dwarfs  for  two  different  reasons.
First, the color criterion  of \iz$>\!1.3$~mag was primarily motivated
by  the $z\!\simeq\!6$ galaxy  surveys and  is $\sim$0.5~mag  too blue
\citep{hawl02} to have included only \LT dwarfs.  When we repeated the
above  analysis for  \iz$\geq$1.8~mag \citep{hawl02}  the  star counts
were reduced by $\sim$50\% and  the inferred vertical scale height was
$300\!\pm\!100$~pc.   While there is  a significant  contribution from
late M-dwarfs, the vertical scale height was unaffected by the 0.5~mag
color difference.  Second, an  appreciably reddened M-dwarf could have
an \iz\ color of an  unreddened L- or T-dwarf.  This scenario requires
considerable  reddening, only the  two fields  with \absb\  have $E\iz
\!\gtrsim\!0.1$~mag.   This effect  could account  these  fields lying
more than $1\sigma$ above  the best-fit line in \fig{surfres}, however
we cannot  be certain without more broad-band  filters or spectroscopy.
We investigated this effect's contribution by removing the two suspect
fields  and repeated  the analysis.   While this  procedure reduced the 
observed star counts by $\sim$50\%, it resulted in  a vertical scale
height  of   360$\pm$180~pc.   Without  further   observations  it  is
difficult  to definitively remove  highly reddened  M-dwarfs, however
their contribution should not grossly affect our main goal.

\section{Discussion} \label{discussion}

Using  a  suite of  Monte  Carlo  simulations  and $\nfields$  HST/ACS
parallel fields, we find a vertical scale height of $\finz$~pc for the
\LT  dwarf  population  based  on  $\nltds$  faint  candidates.   This
estimated  scale  height  is   consistent  with  the  known  trend  of
increasing scale  height with decreasing stellar  mass, independent of
reddening,  color selections,  and  other Galactic  parameters and  is
within the  uncertainties of and  is largely a refinement  of previous
work \citep{liu02,grapes}.   Using our value of the  scale height, and
the  parameters  given  in  \S~\ref{model},  we  predict  a  total  of
$\sim\!10^{11}$     \LT    dwarfs    and     a    total     mass    of
$\lesssim10^{9}$~M$_{\odot}$ in the Milky Way.

This  improved understanding  of the  \LT dwarf  Galactic distribution
will aide high-redshift surveys in better estimating the contamination
of  \LT  dwarfs in  their  samples.  In  the  recent  HUDF pointed  at
$(\alpha,\delta)$=(3$^{\mathrm                          h}$32$^{\mathrm
m}$39$\fs$0,--27\degr47'29$\farcs$1) and a depth of $z'\!\sim$29~mag,
we predict $\gtrsim$2 \LT dwarfs in its $\sim$11~arcmin$^{2}$ field of
view,   which  has   been   confirmed  by   \citet{grapes}  who   have
spectroscopically  identified  three  \LT  dwarfs.   We  confirm  that
Galactic  \LT dwarfs cannot  significantly corrupt  the $z\!\simeq\!6$
surveys  in  high-latitude  fields  (the HUDF  for  example),  however
low-latitude  fields will  find  a modest  number  of interloping  \LT
dwarfs.  With only $\nltds$ candidates from $\nfields$ fields, our 
statistics remain sparse and ideally require further observations.

\acknowledgments This  work was  funded by the  ASU NASA  Space Grant.
The authors  thank Dave  Burstein and Neill Reid for  their helpful
discussions  on Galactic structure  and \LT  dwarfs.  The authors are 
greatly appreciative for the referee's insightful and useful comments.
We wish to dedicate this work to the memory of Dr. John Bahcall.

Facilities:  \facility{HST(ACS)}

\end{document}